\documentclass[12pt]{iopart}
\usepackage{graphicx}
\usepackage[active]{srcltx}
\usepackage{amsfonts}
\usepackage[latin1]{inputenc}
\graphicspath{{FIGURES/}}

\begin{document}
\title[Vertex routing models]{Vertex routing models}

\author{D Markovic$^1$ and C Gros$^2$}
\address{$^1$ Institute for Theoretical Physics, 
             Johann Wolfgang Goethe University, 
            Frankfurt am Main, Germany}
\address{$^2$ Institute for Theoretical Physics, 
             Johann Wolfgang Goethe University, 
            Frankfurt am Main, Germany}
\date{\today}

\begin{abstract}
A class of models describing the flow of information
within networks via routing processes is proposed and investigated,
concentrating on the effects of memory traces on the
global properties. The long-term flow of information is
governed by cyclic attractors, allowing to define a measure 
for the information centrality of a vertex given by the 
number of attractors passing through this vertex. We find
the number of vertices having a non-zero information centrality 
to be extensive/sub-extensive for models with/without
a memory trace in the thermodynamic limit. We
evaluate the distribution of the number of cycles,
of the cycle length and of the maximal basins
of attraction, finding a complete scaling collapse in the
thermodynamic limit for the latter. Possible implications
of our results on the information flow in social networks
are discussed.
\end{abstract}
\pacs{89.75.Hc, 02.10.Ox, 87.23.Ge, 89.75.Fb}
\maketitle

\section{Introduction}
 
The structural and statistical properties 
of evolving and dynamical networks have 
been studied intensively over the last decade 
\cite{Albert02,Dorogovtsev02,boccaletti2006},
due to their ubiquitous importance in technology, 
the realms of life and complex system theory 
in general \cite{Gros08}. Transmission of physical
quantities like electricity and of information are
key network functionalities, both in physical
networks like the power grid and the Internet, as
well as in relational networks such as social networks
\cite{granovetter73}. The basic transmission process takes 
place between two network vertices and two constituent 
vertices are linked by an edge whenever direct transmission 
is possible.

Another key network functionality is routing. An
incoming physical quantity, package or information,
arriving at a certain vertex is forwarded by this vertex.
This routing process may proceed either via static routing
tables or via dynamical routing protocols. The latter is
the case for the Internet, the internet servers having the
task of routing information packages such that they 
find their way eventually to the addressees specified
in the package headers. Here we specify a class of deterministic
vertex routing models with static routing tables, viz
with quenched routing dynamics. The routing tables are 
drawn randomly for every vertex and the models are 
characterized by the network topology on one side and 
by the length of the memory trace along the routing 
path on the other side. 

\begin{figure}[tb]
\centerline{
\includegraphics[width=0.4\textwidth]{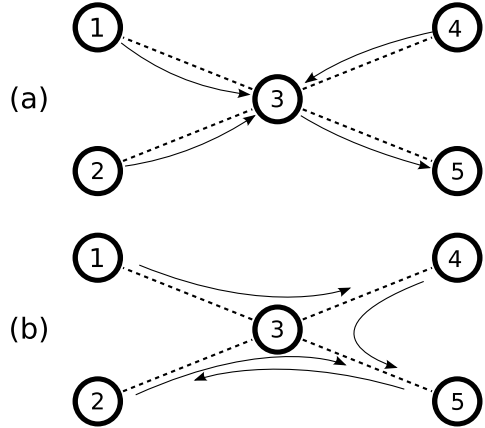}
           }
\caption{Examples of routing tables. (a) Information
is always routed to vertex 5, independently of where it
came from (memoryless model) (b) Information arriving
at vertex 3 from the vertices 1,4,5,2 is routed to the
vertices 4,5,2,5 respectively (model with memory trace).
}
\label{fig_routingIllust} 
\end{figure}

In this study we focus on the effect of the routing memory
on the long-term dynamical properties of the routing process,
considering the case of information routing.
For this purpose we consider fully connected networks 
and two kinds of trace memory. In the first
case memory is absent and the package is passed on 
irrespectively of where it came from, always along the same
outgoing edge, see Fig.\ \ref{fig_routingIllust}.
In the second case the memory trace consists of a single time step,
and the routing of incoming information depends on 
the vertex that routed it in the previous time step;
for every incoming edge the routing table specifies a
distinct outgoing edge. We study then the statistics 
of the resulting cyclic attractors, the basins
of attractions and of a measure for the
degree of information centrality. The vertex routing
models are defined in the phase space of 
directed links and a given vertex is information central 
of degree $c=0,1,2,...$ when it belongs to one or more 
intersecting attractors of the information routing dynamics.

We find that a memory trace for the routing process makes
a qualitative difference. In the absence of memory only
a sub-extensive number $O(N^{1/2})$ of vertices is information 
central in the thermodynamic limit $N\to\infty$. For the case
of a one-timestep memory trace the number of information
central vertices is on the other hand extensive, being
linearly proportional to the number of vertices $N$.

The concept of topological-based centrality and its 
dependence on network properties has been widely 
studied \cite{Albert02,Dorogovtsev02,boccaletti2006}.
The notion of information centrality used here is, on
the other side, based on the observation, that the flux
of information the members of a social network receive
is important. This flux of information 
is maximal whenever a person is part of one or more
attractors of the information routing process, as
it dispose in this case over the entire information 
generated in the respective basins of attraction.
Members of a social network located on the fringe
of the information flow will, on the other side,
receive information only from a small number of other
members. 

We note that the standard network characterization of real-world
networks is provided in terms of network topologies 
\cite{wasserman94}. We propose that there is a need
to supplement the field data with information
describing the dynamics of routings, which would allow
to evaluate the possible social relevance of the information
flux and accumulation. The vertex routing models 
considered here may, in addition, be regarded
as a reference models, akin to the Erd\"os-R\'enyi 
model of graph theory \cite{Erdos59} and to the $NK$-model 
of dynamical boolean networks \cite{Kauffman69},
having well defined and controllable dynamical properties 
in the thermodynamic limit.
\section{Vertex routing models}

Our models
are based on the idealized notion, that
a vertex $V_k$ receiving information from a vertex $V_j$ 
will transmit it to one other vertex only,
say $V_i$. A vertex routing table $\hat T$ then
corresponds to the binary tensor $T_{ijk}=(\hat T)_{ijk}$,
\begin{equation}
T_{ikj}\ =\ \left\{
\begin{array}{rrl}
0 &\quad\hbox{no\ routing}& \\
1 &\quad\hbox{routing\ from} &  
\stackrel{\longrightarrow}{(jk)}
\hbox{\ to\ } \stackrel{\longrightarrow}{(ki)}
\end{array}
            \right. ~,
\label{eq_routing_table}
\end{equation}
where $\stackrel{\longrightarrow}{(jk)}$ denotes
the directed edge from vertex $V_j$ to vertex $V_k$,
compare Fig.\ \ref{fig_routingIllust}.
For every incoming link $\stackrel{\longrightarrow}{(jk)}$
to the vertex $V_k$
the information is routed to a single
outgoing link $\stackrel{\longrightarrow}{(ki)}$, thus
\begin{equation}
\sum_i T_{ikj} \ =\ 1, \qquad 
\sum_{ij} T_{ikj} \ =\ z_k\ \equiv\ N-1~, 
\label{eq_table_sum_rules}
\end{equation}
where $z_k$ is the degree of the routing
vertex $V_k$. Considering here a fully connected
graph with $N$ sites we have $z_k\equiv N-1$.
The entries $T_{ikj}=0,1$ of the routing table 
are determined consecutively for all vertices: 
For a given vertex $k$ and a given 
incoming edge $\stackrel{\longrightarrow}{(jk)}$ one
outgoing edge $\stackrel{\longrightarrow}{(kI)}$
is randomly chosen among the $N-2$ potential candidates
of outgoing edges. Then $T_{Ikj}=1$ and $T_{ikj}=0,\ \forall i\ne I$.
For the two models we make the following differentiation:
\begin{itemize}
\item {\sl Without memory} For every directed edge
      the entries $T_{ikj}$ are selected randomly,
      and are independent of the originating vertex, 
thus we can write $T_{ikj}\equiv T_{ikl}$, $\forall i,j,k,l$. 
\item {\sl With memory} Again all entries
      $T_{ikj}$ are drawn randomly. Routing of
      information depends on where it came from, but
      backrouting is not allowed: $T_{jkj}\equiv0$,
      $\forall k,j$.
\end{itemize}
%

\begin{figure}[t]
\centerline{
\includegraphics[width=0.5\textwidth]{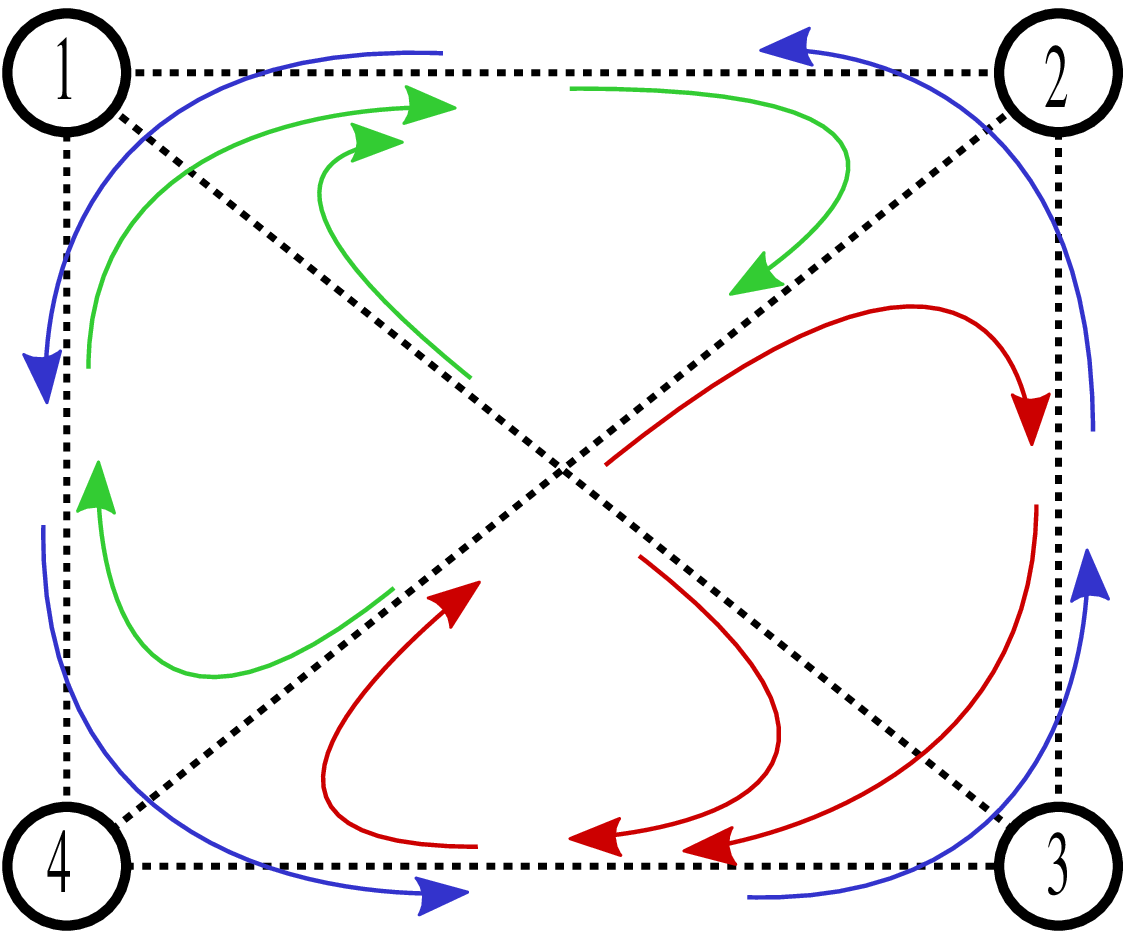}
           }
\caption{
Example of a routing table on a fully connected
network with four nodes. There are three distinct
cyclic attractors containing the directed links $\big[\overrightarrow{(14)},\ 
\overrightarrow{(43)},\ \overrightarrow{(32)},\ \overrightarrow{(21)}\Big]$, 
$\Big[\overrightarrow{(34)},\ \overrightarrow{(42)},\ \overrightarrow{(23)}\Big]$
and
$\Big[\overrightarrow{(12)},\ \overrightarrow{(24)},\ \overrightarrow{(41)}\Big]$
respectively. For the nodes 1 and 3 the information centrality is $c=2$, 
while for nodes 2 and 4 one finds $c=3$.
}\label{fig_examp_cycles} 
\end{figure}

The rational for these two models in the context of
social networks is the following: For the memoryless
case a new information is passed on always to the best
friend, irrespectively of the information source. In the
model with memory the information routing depends on the
source. Information received by a relative might be passed on
to another relative and work-related news might be passed on
predominantly to a work-place buddy. The suppression of
backrouting is not important for large $N$ but clearly makes
sense; It is never a good idea to echo a joke to 
the person which told it in the first place.

An example of a routing table on a fully connected 
graph with four nodes is presented in Fig.~\ref{fig_examp_cycles}. 
In a discrete phase space built upon the directed edges 
every cycle is an attractor, thus we have cyclic 
attractors. In this example we have three cyclic attractors 
(labeled with colors), each of which has a basin of attraction 
of volume $V=4$, which is the number of directed edges. The
states $\overrightarrow{(13)}$ and $\overrightarrow{(31)}$ in
the phase space of the dynamics belong to the basin of attractions of 
the red and the green attractors, respectively, but they do not belong to any
cyclic attractor. Also, for nodes 1 and 3 the information centrality $c=2$, 
while for nodes 2 and 4, $c=3$.

We also note that the vertex routing models
(\ref{eq_routing_table}) play an important role
in the context of neural cognitive information processing.
In this context a vertex corresponds to an object
and the sequence of vertices activated by the
routing process to an associative thought process
\cite{gros07,gros09}. In addition there is a close
relation to random boolean networks 
\cite{aldana03,drossel08}, with the directed links
constituting the boolean variables. In terms of
a boolean network the routing model operate in
the sparse activity limit, since the routing 
problem deals with routing of individual packages,
a single directed link being operative at any
given time.

\begin{figure}[t]
\centerline{
\includegraphics[width=0.5\textwidth]{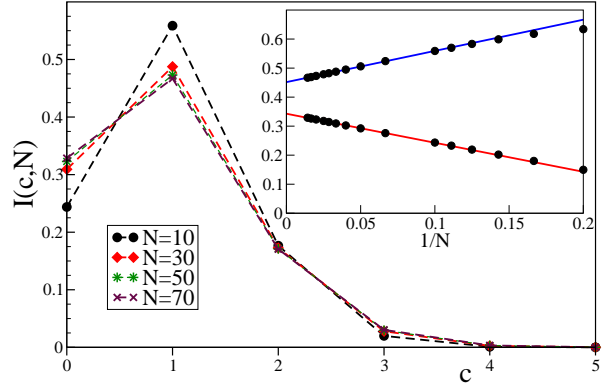}
           }
\caption{
The information centrality $I(c,N)$ for the models with
memory, viz the probability that $c$ attractors pass through
a given vertex, for $N=10,20,\dots,70$ number of vertices
(top-down/down-top for $c=1$/$c=0$).
Inset: $I(0,N)$ (red line) and $I(1,N)$ (blue line) as a function of $1/N$,
       the lines are linear fits.
}
\label{fig_info_centrality} 
\end{figure}

\section{Memoryless model}

In this case
the routing tensor $T_{ikj}$ is independent
of the last index and its dimension is effectively
reduced to two. The probability distribution
$N_l(L,N)$ of finding a cycle of length $L$
in a network with $N$ nodes is given by [see \ref{appenA}]
\begin{equation}
N_l(L,N) \ =\ {1\over z(N)} \,
{N\choose L} {L!\over L }
{1\over (N-1)^L}~,
\label{eq_no_mem_rho_L_N}
\end{equation}
where $z(N)$ is a normalization factor,
${N\choose L}$ the number of $L$ sites
out of the $N$ vertices and ${L!/L}$
the number of possibilities to connect
$L$ sites into distinct loops. The factor
$(N-1)^{-L}$ in (\ref{eq_no_mem_rho_L_N}) 
is the probability that any ordered set of
$L$ sites is connected via the routing
table, with $1/(N-1)$ being the probability
of two vertices being connected. 

For the memoryless model the phase space reduces
effectively to the number of vertices $N$,
compare Fig.~\ref{fig_routingIllust}, and
the cycles are disjoint and non-intersecting,
only $c=0,1$ cycles may pass through any vertex in
the memoryless model; any given vertex may
have only an information centrality $c=0,1$.
Defining by $I(c,N)$ the distribution of 
the information centrality we find [see \ref{appenB}]
$$
I(1,N) = {z(N)\over N}\langle L \rangle
 = \sum_{L=2}^N {(N-1)!\over (N-L)! (N-1)^L}
$$
for the memoryless model,
where we have used (\ref{eq_no_mem_rho_L_N}) and
$\langle L \rangle = \sum_L LN_l(L,N)$.
$I(1,N)$ scales like
\begin{eqnarray}
\label{eq_no_mem_I_1_N}
I(1,N)\ \sim\ 1/\sqrt{N},
\qquad\quad N\to\infty 
\end{eqnarray}
in the thermodynamic limit. Only
a sub-extensive number of vertices belong to
an attractor of the information flow and
essentially all vertices are excluded from the
long-term flow, $I(0,N)\to1$ for $N\to\infty$.

\begin{figure}[tb]
\centerline{
\includegraphics[width=0.5\textwidth]{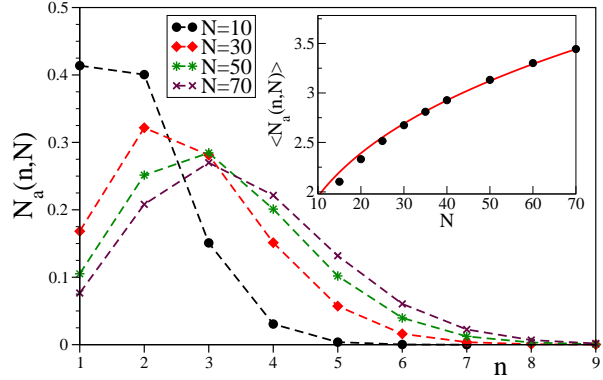}
           }
\caption{
The distribution $N_a(n,N)$ of the total number $n$ of cycles
per realization, for $N=10,30,50,70$ numbers of vertices.
Inset: The average $\langle N_a(n,N)\rangle$, 
as a function of $N$, together with a power-law fit (solid line).
}
\label{fig_number_cycles} 
\end{figure}

\section{Model with memory trace}

We have
performed extensive numerical simulations of the
vertex routing model with a memory trace. We studied 
ensembles of $N_{real}=10^6$ randomly drawn realizations
of the routing tensor, considering fully connected networks
having $N=10,...,70$ sites, and performing an exhaustive
search of all attractors and their respective basins
of attractions. We evaluated the normalized distributions
\begin{equation}
I(c,N), \quad N_a(n,N),\quad N_l(L,N), \quad
V_b^{max}(v,N)~,
\label{eq_defs}
\end{equation}
where $I(c,N)$ is the distribution of the information 
          centrality, with $c$ being the number of attractors 
          passing through a vertex,
$N_a(n,N)$, the distribution of the number
          of attractors $n$ per network,  
$N_l(L,N)$ the distribution of the lengths $L$
          of attractors and the distribution of
the volume of the largest basin of 
attraction $V_b^{max}(v,N)$. For every
model realization all basins of attraction 
were evaluated and the statistics of the
maximal volume is given by $V_b^{max}(v,N)$, with 
the volume $v$ being defined relative to the
entire phase space $\Omega=N(N-1)$ of directed links.
 
We find, see Fig.~\ref{fig_info_centrality}, that
the information centrality approaches a well defined
limiting function $I(c)=\lim_{N\to\infty}I(c,N)$ 
in the thermodynamic limit. The availability of
information is quite democratically distributed,
only a fraction $I(0)\approx0.34$ of vertices 
are cut-off completely from the long term information flow. 

In Fig.~\ref{fig_number_cycles} the distribution
of the number of attractors per network is presented.
The average number of attractors per network 
increases only slowly with the size of the network, as
$\langle N_a(n,N)\rangle \sim N^{\alpha_{a}}$. 
Somewhat larger system sizes are necessary for a 
reliable estimate of the scaling exponent, our
best fit (given in the inset of Fig.~\ref{fig_number_cycles})
indicates $\alpha_{a}\approx0.29$. 

\begin{figure}[tb]
\centerline{
\includegraphics[width=0.5\textwidth]{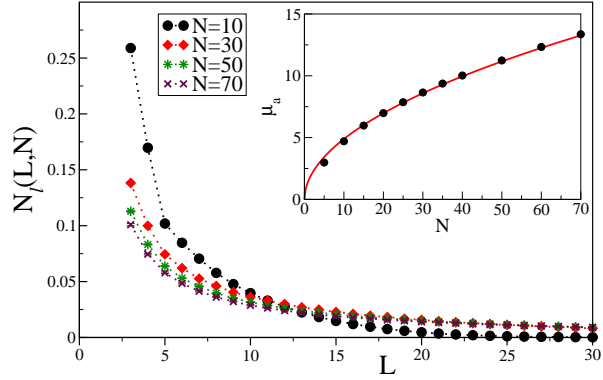}
           }
\caption{
The cycle-length distribution $N_l(L,N)$ 
as a function of cycle length $L$,
for $N=10,20,\dots,70$ number of vertices.
Inset: The median $\mu_a(N)$ of $N_l(L,N)$, as a function of $N$.
}
\label{fig_length_cycles} 
\end{figure}

We encountered problems of undersampling of the space of
all possible model realizations when evaluating
the cycle-length distribution $N_l(L,N)$, presented
in Fig.\ \ref{fig_length_cycles}, which is a phenomena 
well known in the field of random boolean networks \cite{drossel08}.
For the case of the vertex routing model the probability
of finding very long cycles could not be determined
accurately, due to the fat tails of $N_L(l,N)$.
This problem affects also the results for
the mean cycle length $\langle N_l(L,N) \rangle$, but
not the median $\mu_a(N)$ of $N_l(L,N)$. We found scaling
close to a square-root law for the median 
(inset of Fig.\ \ref{fig_length_cycles}), 
$\mu_a(N)\sim N^{0.51}$. 

The number of cyclic attractors steadily increases with $N$,
as shown in Fig.~\ref{fig_number_cycles}. The question
is then, whether there is typically a single dominating attractor,
in terms of the size of the respective basins of
attractions, or whether the phase-space volume 
is more or less equally divided between the attractors
being present. This information 
is provided by $V_b^{max}(v,N)$. The probability that
the largest attractor volume is in the interval $[v_1,v_2]$ is
given by $\sum_{v=v_1}^{v_2} V_b^{max}(v,N)$.
Here the volume is relative to the maximal basin
of attraction, which is equal to the phase-space
volume $\Omega=N(N-1)$ and $v=1$ occurs when only 
a single attractor is present.

\begin{figure}[tb]
\centerline{
\includegraphics[width=0.5\textwidth]{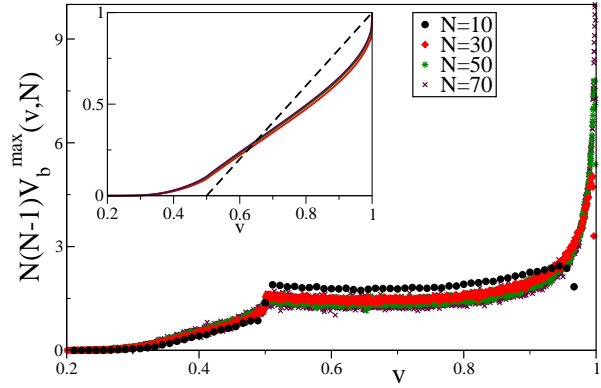}
	}
\caption{
The rescaled distribution $N(N-1)V_b^{max}(v,N)$ of the volumes 
of the basin of attractions, for $N=10,30,50,70$. The volume $v$
is normalized with respective to the total phase space
$\Omega=N(N-1)$ of directed links.
Inset: The integrated distribution $\int_0^v dv'V_b^{max}(v',N)$, 
for $N=30,50,70$ (the curves fall on top of each other). For
comparison the result (\ref{eq_two_random_basins})
for the case of two attractors with randomly 
distributed basins of attractions is given
(dashed line).
}
\label{fig_basin_attraction} 
\end{figure}

The rescaled $\Omega V_b^{max}(v,N)$ converges
rapidly with $N$ to a limiting function, see
Fig.\ \ref{fig_basin_attraction}. 
There is a divergence for $v\to1$, due to the fact that
the probability of finding an ensemble realization
with a single attractor, $N_a(1,N)$, scales like
$1/N$ and consequently
$\Omega V_b^{max}(v=1,N)=\Omega N_a(n=1,N)\sim N$.
A kink at $v=1/2$ occurs for $V_b^{max}(v,N)$. For
$v>1/2$ ensemble realizations containing two or more
cycles contribute to $V_b^{max}(v,N)$, for
$v<1/2$ realizations with three or more attractors
contribute.

The divergence of $\Omega V_b^{max}(v,N)$
for $v\to1$ seems to indicate that cycles with
large volumes of attraction would have a dominating
role, controlling most of the long-term information flow.
In order to understand the origin of this divergence
we have compared the 
integrated distribution $\int_0^v dv' V_b^{max}(v',N)$ 
(inset of Fig.\ \ref{fig_basin_attraction})
with an unbiased distribution of basins of attraction,
taking the case of two attractors (dashed line in
the inset of Fig.\ \ref{fig_basin_attraction}). In
this case, when only two attractors are present, one of the volumes
of attraction is always equal or larger than $\Omega/2$,
and 
\begin{equation}
\int_0^v dv'V_b^{max}(v',N)\ =\
\left\{
\begin{array}{rcl}
 0 &\  & v<1/2 \\
2v-1 &\  & v>1/2
\end{array}
\right. ~.
\label{eq_two_random_basins}
\end{equation}

The integrated distribution of model with a memory trace
follows somewhat the simplified model 
(\ref{eq_two_random_basins}) of a random distribution 
of two basins of attraction, albeit with a substantial 
suppression close to unity, indicating that the 
divergence of $\Omega V_b^{max}(v,N)$
for $v \to 1$ is a directly related to the statistical
properties of the distribution of basins of attraction,
independent of the details of the dynamics.

\section{Discussion}

We have presented and analyzed a novel
class of models suitable for describing
the flow of information in complex networks.
In these models the flow of information is
realized by information packages travelling
along the edges of a communication network,
whereas it is assumed, in the intensively studied
information diffusion models, that information is 
an attribute attached to the vertices and not 
to the edges. The dynamics is determined in
our class of models by routing tables, specifying 
respectively for every vertex the routing
of incoming information packages. 

The two vertex routing models discussed in 
the present study constitute idealized reference
models, not suitable for a direct
modeling of field data. The key point
of the present study is to analyze the
non-trivial effects of a memory trace
on the long-term information flow and
not the dependence on the network topology,
which is left for further studies.
Information is conserved in the present
models. This would be clearly not the case for
real-world social networks but it holds 
for the network of internet routers, which have 
the task of routing packages of information without
increasing or decreasing their number.

Traditionally, the flow of information in social
networks, like the random spreading of rumors, has been 
modeled by diffusion processes \cite{cointet07}.
Searching for more realistic models the special topology 
of social networks has been discussed intensively \cite{newman03},
as well as corrections to the diffusion process itself
\cite{wu04}. The vertex routing and the diffusion models 
of information flow constitute two extremes. In the
first model the direction of the information flow 
is 100\% deterministic, in the second model 100\% random.
The information-flow occurring in real-world social nets is 
expected to be partially a random and partially a directed
process. It will therefore be of interest for future studies
to interpolate between these two reference models. We
propose that field studies characterizing social networks
should be supplemented by data describing the dynamics of
the information flow, in addition to the standard 
structural and topological characterization, as the
accumulation of information in the attractors of the
information flow may have a substantial social impact.

The dependence of the routing dynamics on the network
topology is additionally an important issue for future studies.
Cycles in the routing process can only then appear,
when the underlying network topology allows for loops.
Here we have been considering fully connected networks and
loops of all length are present. For many classes of
real-world networks there is a characteristic loop-length,
which generally scales sub-extensively with the network-size
\cite{rozenfeld05}. Interesting interferences phenomena between 
this scaling and the sub-extensive scaling of the typical
attractor-length (inset of Fig.\ \ref{fig_length_cycles})
may then be expected.

\appendix
\section{The cycle length distribution for the memoryless model} \label{appenA}
Let $q_t$ be the probability that a path remains unclosed
after t steps. If a path is still open at time $t$, we have
already visited $t+1$ different nodes. There are $t$ ways to
close the path in the next time step. The relative probability
is then $\rho_t = t/(N-1)$. The probability of still having an
open path after $t$ steps is
\begin{displaymath}
 q_{t+1}=q_t(1-\rho_t)=\prod_{i=0}^t \Big(\frac{N-i-1}{N-1}\Big),\qquad q_0=1
\end{displaymath}
\begin{displaymath}
 \Rightarrow\qquad q_t = \frac{(N-1)!}{(N-1)^t(N-1-t)!}~.
\end{displaymath}
The average number of cycles of length $L$ is
\begin{displaymath}
 \langle N_l(L)\rangle = \frac{q_{t=L-1}}{N-1}\frac{N}{L}=\frac{N!}{L(N-1)^L(N-L)!}~,
\end{displaymath}
where we used the following considerations \cite{Gros08}:
\begin{enumerate}
 \item The probability that the node, visited at time $t+1$, is
identical to the starting node is $1/(N-1)$.
\item There are $N$ possible starting points.
\item Factor $1/L$ corrects for the overcounting of cycles when
considering the L possible starting sites of the L-cycle.
\end{enumerate}
After normalization we obtain Eq.$~$(\ref{eq_no_mem_rho_L_N}) , the probability
distribution of cycle lengths
\begin{displaymath}
 N_l(L,N) = \frac{1}{z(N)}\frac{N!}{L(N-1)^L(N-L)!}~,
\end{displaymath}
where $z(N)$ is normalization factor
\begin{displaymath}
z(N) = \sum_{L=2}^N \frac{N!}{L(N-1)^L(N-L)!}
= \frac{N!}{(N-1)^N}\sum_{k=0}^{N-2} \frac{(N-1)^k}{(N-k)k!}~.
\end{displaymath}

\section{Connection between the average information centrality
and the average cycle length}\label{appenB}

We consider an ensemble of $R$ random realizations of
a routing tensor on a fully connected network with $N$ nodes.
Let $n_\alpha$ be the total number of vertices which belong
to at least one cyclic attractor, where $\alpha = 1\ldots R~$.

The length of an cyclic attractor is equivalent to the number of vertices
that belong to the attractor. In the case of only one existing attractor
$n_\alpha = L-\sum_{r=1}^{r_{max}}(r-1)Q_\alpha(r,N)$, where L is the length 
of the attractor and $Q_\alpha(r,N)$ is the number of nodes which are repeated $r$ 
times during one cycle of the cyclic attractor.
If we denote the number of cycles of length $L$ with
$N_\alpha(L,N)$, we can write, for the case of more than
one co-existing cyclic attractors, the following relation
\begin{equation}\label{B_first}
 n_\alpha = \sum_L LN_\alpha(L,N)-\sum_{r=1}^{r_{max}}(r-1)Q_\alpha(r,N)-\sum_{c=1}^{c_{max}}(c-1)P_\alpha(c,N)~,
\end{equation}
where $P_\alpha(c,N)$ is number of nodes with information
centrality $c$. For example let us consider the case
when we have two attractors \big[$\overrightarrow{(12)}$,
 $\overrightarrow{(23)}$, $\overrightarrow{(31)}$ \big] 
and \big[$\overrightarrow{(14)}$, $\overrightarrow{(45)}$, $\overrightarrow{(51)}$, 
$\overrightarrow{(16)}$, $\overrightarrow{(67)}$, $\overrightarrow{(71)}$\big].
It is easy to count that there are 7 distinct vertices
contained in this two attractors. We see that only node 
1 has information centrality c=2, thus
$P_\alpha(c=1) = 6$ and $P_\alpha(c = 2) =1$. Also, node
1 is the only one two repeat two times during one cycle of
second attractor, thus $Q_\alpha(r=1) = 6$ and $Q_\alpha(r=2)=1~$.
If we put this values into (\ref{B_first}) we obtain $n_\alpha = 7$.

On the other hand, if we denote the distribution presented in
Fig.~\ref{fig_info_centrality} as $I(c,N)$, where $c$ is
the information centrality, then, in the case of one
given realization of the routing tensor, we can write
\begin{equation}\label{B_second}
 P_\alpha(c,N) = N\cdot I_\alpha(c,N)~.
\end{equation}
Also, total number of vertices, which are members of
cyclic attractors, is then
\begin{equation}\label{B_third}
 n_\alpha = N\sum_{c=1}^{c_{max}}I_\alpha(c,N)~.
\end{equation}

Combining (\ref{B_first}) and (\ref{B_second}) with (\ref{B_third})
we obtain
\begin{eqnarray}\label{B_fourth}
N\sum_{c=1}^{c_{max}}I_\alpha(c,N) = &
\sum_L LN_\alpha(L,N)-N\sum_{c=1}^{c_{max}}(c-1)I_\alpha(c,N)-\\
& -\sum_{r=1}^{r_{max}}(r-1)Q_\alpha(r,N)~, \nonumber
\end{eqnarray}

After averaging over entire ensemble of $R$ realizations
of routing tensor and dividing both sides of (\ref{B_fourth}) with $N$, we obtain
\begin{equation}\label{B_fifth}
 \sum_{c=1}^{c_{max}}cI(c,N)=\frac{z(N)}{N}\sum_L LN_l(L,N)-\frac{1}{N}\sum_{r=1}^{r_{max}}(r-1)Q(r,N)~,
\end{equation}
where we have used following relations
\begin{eqnarray*}
\frac{1}{R}\sum_{\alpha=1}^R N_\alpha(L,N) = z(N) \cdot N_l(L,N)~, \\
\frac{1}{R}\sum_{\alpha=1}^R I_\alpha(c,N) = I(c,N)~,\\
\frac{1}{R}\sum_{\alpha=1}^R Q_\alpha(c,N) = Q(c,N)~.
\end{eqnarray*}

In the case of the memoryless model, $I(c,N)$ have non zero
values only for $c=0,1$, and it is not possible that one node
is repeated more than once in one cycle of cyclic attractor, thus
$\frac{1}{N}\sum_r(r-1)Q(r,N) = 0$. Therefore, from 
(\ref{B_fifth}) we obtain
\begin{equation}
I(1,N)=\frac{z(N)}{N}\langle L\rangle~,
\end{equation}
where $\langle L\rangle = \sum_LLN_l(L,N)~$, which is the central
result of this appendix.



\section*{References}
\begin{thebibliography}{99}

\bibitem{Albert02} R.~Albert and A.~Barab\`{a}si, 
{\it Statistical mechanics of complex networks}, 
Reviews of Modern Physics {\bf 74}, 47 (2002).
%
\bibitem{Dorogovtsev02} S.N.~Dorogovtsev and J.F.F.~Mendes,
{\it Evolution of networks}
Advances in Physics {\bf 51}, 1079 (2002).
%
\bibitem{boccaletti2006} S. Boccaletti, V. Latora, Y. Moreno, 
          M. Chavez  and D.U.Hwang, 
{\it Complex networks: Structure and dynamics},
Physics Reports {\bf 424}, 175 (2006).
%
\bibitem{Gros08} C. Gros,
{\it Complex and Adaptive Dynamical Systems, A Primer},
              Springer (2008).
%
\bibitem{granovetter73} M.S. Granovetter,
{\it The Strength of Weak Ties},
American Journal of Sociology {\bf 78}, 1360 (1973).
%
\bibitem{wasserman94} S. Wasserman and K. Faust, 
{\it Social Network Analysis: Methods and Applications},
Cambridge University Press (1994).
%
\bibitem{Erdos59} P.~Erd\"os and A.~R\'enyi,
{\it On random graphs}
Publicationes Mathematicae {\bf 6}, 290 (1959).
%
\bibitem{Kauffman69} S.A. Kauffman,
{\it Metabolic Stability and Epigenesis in Randomly
       Constructed Nets},
Journal of Theoretical Biology {\bf 22}, 437 (1969).

\bibitem{gros07} C. Gros,
{\it Neural networks with transient state dynamics}, 
New Journal of Physics {\bf 9}, 109 (2007).

\bibitem{gros09} C. Gros,
{\it Cognitive computation with autonomously active neural networks:
       an emerging field},
Cognitive Computation {\bf 1}, 77 (2009).

\bibitem{aldana03} M. Aldana-Gonzalez, S. Coppersmith and L.P. Kadanoff, 
{\it Boolean Dynamics with Random Couplings},
in E. Kaplan, J.E. Marsden, K.R. Sreenivasan (eds.),
Perspectives and Problems in Nonlinear Science,
Springer (2003).
 
\bibitem{drossel08} B. Drossel,
{\it Random Boolean Networks},
in H.G. Schuster (ed), Reviews of Nonlinear 
Dynamics and Complexity {\bf 1} (2008).

\bibitem{cointet07} J.P. Cointet, and C. Roth, 
{\it How Realistic Should Knowledge Diffusion Models Be?},
Journal of Artificial Societies and Social Simulation
{\bf 10}, 3 (2007).

\bibitem{newman03} M.E.J. Newman, 
{\it The Structure and Function of Complex Networks},
SIAM Review {\bf 45}, 167 (2003).

\bibitem{wu04} F. Wu, B.A. Huberman, L.A. Adamic, and J.R. Tyler, 
{\it Information flow in social groups},
Physica A {\bf 337}, 1 (2004).

\bibitem{rozenfeld05} H.D. Rozenfeld, J.E. Kirk, E.M. Bollt, and D. Ben-Avraham, 
{\it Statistics of cycles: How loopy is your network?}, 
J.Phys. A: Math. Gen. {\bf 38}, 4589-4595 (2005)
%
%
%
%
\end{thebibliography}
\end{document}